\documentclass[aps,prd,twocolumn,nofootinbib,superscriptaddress,preprintnumbers,floatfix]{revtex4-1}
\usepackage{amsfonts,amsmath,amssymb,amsthm,braket,nicefrac}
\usepackage{booktabs,array}
\usepackage[utf8]{inputenc}
\usepackage{float,graphicx,hhline}
\usepackage{appendix}
\usepackage[dvipsnames]{xcolor}
\usepackage{mathtools}
\usepackage{mathrsfs}

\usepackage{hyperref}
\hypersetup{
    pdftitle={},
    colorlinks=true,     
    linkcolor=blue,      
    citecolor=blue,      
    filecolor=blue,      
    urlcolor=blue        
}
\usepackage[nameinlink]{cleveref}

\newcommand{\xx}{\mathbf{x}}

\newcommand{\Tr}{\mathrm{Tr}\,}
\newcommand{\RE}{\mathrm{Re}\,}

\newcommand{\dd}{\mathrm{d}}
\newcommand{\EE}{\mathbf{E}}

\newcommand{\rr}{\mathbb{R}}

\newcommand{\BB}{\mathbf{B}}
\newcommand{\Pcal}{\mathcal{P}}

\newcommand{\Ccal}{\mathcal{C}}

\newcommand{\SU}{\mathrm{SU}}

\newcommand{\flow}{\mathrm{flow}}
\newcommand{\Pscr}{\mathscr{P}}
\newcommand{\Dscr}{\mathscr{D}}

\graphicspath{{./}}

\newcommand{\getMPIAffiliation}{\affiliation{Max Planck Institute for Mathematics in the Sciences, Inselstraße 22, Leipzig, 04103, Germany}}
\newcommand{\getHeidelbergAffiliation}{\affiliation{Institute for Theoretical Physics, Heidelberg University, Philosophenweg 16, 69120 Heidelberg, Germany}}
\newcommand{\getEMMIAffiliation}{\affiliation{ExtreMe Matter Institute EMMI, GSI, Planckstr.~1, 64291 Darmstadt, Germany}}
\newcommand{\getMITAffiliation}{\affiliation{Center for Theoretical Physics, Massachusetts Institute of Technology, Cambridge, MA 02139, USA}}
\newcommand{\getIAIFIAffiliation}{\affiliation{The NSF AI Institute for Artificial Intelligence and Fundamental Interactions}}

\begin{document}

\title{Topological data analysis of the deconfinement transition in SU(3) lattice gauge theory}

\author{Daniel Spitz}
\email{daniel.spitz@mis.mpg.de}
\getMPIAffiliation

\author{Julian M. Urban}
\getMITAffiliation
\getIAIFIAffiliation

\author{Jan M. Pawlowski}
\getHeidelbergAffiliation
\getEMMIAffiliation

\preprint{MIT-CTP/5801}

\begin{abstract}
We study the confining and deconfining phases of pure $\SU(3)$ lattice gauge theory with topological data analysis.
This provides unique insights into long range correlations of field configurations across the confinement-deconfinement transition.
Specifically, we analyze non-trivial structures in electric and magnetic field energy densities as well as Polyakov loop traces and a Polyakov loop-based variant of the topological density. 
The Betti curves for filtrations based on the electric and magnetic field energy densities reveal signals of electromagnetic dualities.
These dualities can be associated with an interchange in the roles of local lumps of electric and magnetic energy densities around the phase transition. 
Moreover, we show that plaquette susceptibilities can manifest in the geometric features captured by the Betti curves.
We also compare these findings against earlier results for $\SU(2)$ and elaborate on the significant differences. 
Our results demonstrate that topological data analysis can identify clear differences between phase transitions of first and second order for non-Abelian lattice gauge theories and provides unprecedented insights into the relevant structures in their vicinity.
\end{abstract}

\maketitle

\section{Introduction}

Understanding the dynamical mechanism responsible for confinement in non-Abelian gauge theories remains an outstanding challenge. 
Crucially, the deconfinement phase transition is of second order for only a few simple gauge groups while being first order in the majority of cases, at least in four space-time dimensions~\cite{Holland:2003kg, Holland:2003jy, Pepe:2006er}. 
A unifying property among the various proposed confinement mechanisms is the occurrence of topological configurations or defects. 
These configurations are typically drowned in short range fluctuations, and only become visible via cooling. 
The latter, however, changes the underlying physics and complicates the access to the underlying dynamics. 

In recent years, topological data analysis (TDA) has emerged as a promising tool to robustly identify and study geometric objects of varying shapes in lattice data.
Persistent homology---the prevailing TDA method---allows for the identification of topological features along with measures of their dominance, sweeping through a hierarchy of topological spaces inferred from the data~\cite{otter2017roadmap, ChazalMichelIntro}.
For pure $\SU(2)$ lattice gauge theory in particular, persistent homology has been demonstrated to allow for a comprehensive picture of confining and deconfining phases~\cite{Spitz:2022tul}.
Specifically, it was shown that topological densities form spatio-temporal lumps, along with signals of the classical probability distribution of instanton-dyons and its temperature dependence.
Importantly, this analysis is not particularly biased towards detecting certain topological objects defined \emph{a priori}, but is designed to reveal relevant structures in a completely data-driven approach.
In the context of lattice gauge theory, persistent homology has also been employed to probe strings, center vortices, and monopoles~\cite{Sehayek:2022lxf, Sale:2022qfn, Crean:2024nro}.
Moreover, the method has been shown to be sensitive to the intricate phase structures of various condensed matter and statistical systems~\cite{Hirakida:2018bkf, speidel2018topological,santos2019topological, olsthoorn2020finding,tran2021topological, cole2021quantitative, tirelli2021learning, Olsthoorn:2021ems, Kashiwa:2021ctc, sale2022quantitative,he2022persistent, Cao:2022hre, salvalaglio2024persistent}.

In the present work we explore the confining and deconfining phases of pure $\SU(3)$ lattice gauge theory via persistent homology, an important step towards investigations in full quantum chromodynamics (QCD). 
We utilize cubical complexes for different gauge-invariant sublevel set filtrations of the lattice data and investigate their dependence on the gauge coupling, corresponding to different effective temperatures.
Specifically, we focus on electric and magnetic field energy densities as well as local Polyakov loop traces and a Polyakov loop-based topological density, at times comparing with results obtained after cooling/smoothing the raw field configurations.
Contrasting our findings with earlier insights for $\SU(2)$~\cite{Spitz:2022tul}, this physics-informed approach reveals qualitative differences between the two theories and the nature of their phase transitions.
Excitingly, here we are able to identify signals reminiscent of electromagnetic dualities in the vicinity of the phase transition, as well as a structural equipartition at the transition point, where local lumps of electric and magnetic energy densities appear to interchange their roles.
Plaquette susceptibilities are also shown to manifest in the Betti curves through finite-volume effects.
Furthermore, the Polyakov loop-based filtrations appear barely sensitive to the phase transition, in contrast to the situation encountered earlier with gauge group $\SU(2)$.

This paper is structured as follows. In \Cref{Sec:Background} we provide some details on the lattice setup and a brief discussion on common descriptions of the deconfinement phase transition in pure $\SU(3)$ lattice gauge theory. Persistent homology is introduced in 
\Cref{Sec:SelfDualExcitationsBettiElMagn}, and its application to the electric and magnetic energy density filtrations is discussed. 
In \Cref{Sec:Polyakov} we discuss results for Polyakov loop-based filtrations. A short summary of the results and an outlook is provided in  \Cref{Sec:Conclusions}.

\section{Background}\label{Sec:Background}

\subsection{Lattice setup}\label{Sec:LatticeSetup}

We study pure $\SU(3)$ gauge theory discretized on a four-dimensional Euclidean lattice of size $V = N_\sigma^3\times N_\tau$ with periodic boundary conditions in all directions.
Focusing on qualitative, phenomenological aspects of observables based on TDA near the first-order deconfinement transition, we choose $N_\sigma=32$ and $N_\tau=12$ throughout this work and postpone a detailed investigation of the dependence on the particular choice of lattice geometry to the future.
For later use, the set of all lattice sites is denoted~$\Lambda$ and the set of all spatial sites $\Lambda_\sigma$.

Gluonic degrees of freedom are described by means of link variables $U_\mu(x)$ for $x\in \Lambda$, $\mu=1,\ldots,4$, that form $\SU(3)$ group elements.
We employ the conventional Wilson action,
\begin{equation}\label{Eq:Action}
    S[U] = \frac{\beta}{3}\sum_{P}\RE\Tr[1-U_P]\,,
\end{equation}
where $\beta$ is proportional to the inverse squared gauge coupling and the sum runs over all plaquette variables on the lattice, defined as 
\begin{equation}
    {U_P \equiv U_{\mu\nu}(x)=U_\mu(x)U_\nu(x+\hat{\mu})U_\mu^\dagger(x+\hat{\nu})U^\dagger_\nu(x)}
\end{equation}
for an elementary square in the $\mu$-$\nu$ plane at $x\in \Lambda$.
We study the system for $\beta$ ranging from 4.0 to 8.5, which corresponds to increasing the  temperature of the system.

Configurations are generated using the well-known \mbox{(pseudo-)}heatbath algorithm for $\SU(3)$ combined with overrelaxation~\cite{Creutz:1980zw, Cabibbo:1982zn, Kennedy:1985nu, Brown:1987rra, Adler:1987ce}.
In our setup, advancing the Markov chain by 1 step corresponds to 1 heatbath sweep followed by 5 overrelaxation sweeps.
We thermalize each process with 1000 steps using a warm start where all links are sampled uniformly from the Haar measure, and we discard 200 steps between recorded samples to minimize effects of autocorrelation.
Expectation values of persistent homology observables are computed as ensemble averages over 80 samples from 8 independent Markov chains for each $\beta$.
Furthermore, some additional reference results for certain observables are obtained from measurements performed every 10 steps, also using 8 parallel chains per value of $\beta$ and a total of 600 steps, resulting in a total of 480 samples per run.
In particular, we use this second ensemble to compute the volume average of the absolute Polyakov loop trace, as well as the standard volume-scaled two-point susceptibility for several quantities.
For some observable $\mathcal{O}$, the latter is defined as
\begin{equation}\label{Eq:Susceptibility}
    \chi_\mathcal{O} = V \cdot \left(\langle \mathcal{O}^2 \rangle - \langle \mathcal{O} \rangle^2 \right)\,.
\end{equation}
In general, results are given in lattice units, and no scale setting is performed at this stage.
Errors for all results are estimated using the statistical jackknife method; see \Cref{App:JackknifeError} for details.
For the determination of susceptibilities, we perform binning of measurements with a bin size of 10 before computing the associated errors.

As in our previous work~\cite{Spitz:2022tul}, we also compare observables computed from the raw data to the same results obtained from smoothed gauge configurations, from now on referred to respectively as `(un-)cooled' results.
This provides further insight into whether cooling is generally required in order to expose the infrared physics under investigation.
To this end, we employ the Wilson flow~\cite{Luscher:2010iy} based on the Wilson action defined above and using a standard fourth-order Runge-Kutta discretization scheme with a step size of $\delta t = 0.01$ and varying total flow times.
Specifically, some persistent homology results obtained from the uncooled data are re-computed after $n_\flow \in \{1,2,5,10,20\}$ steps.

\begin{figure}
    \centering
	\includegraphics[scale = 0.82]{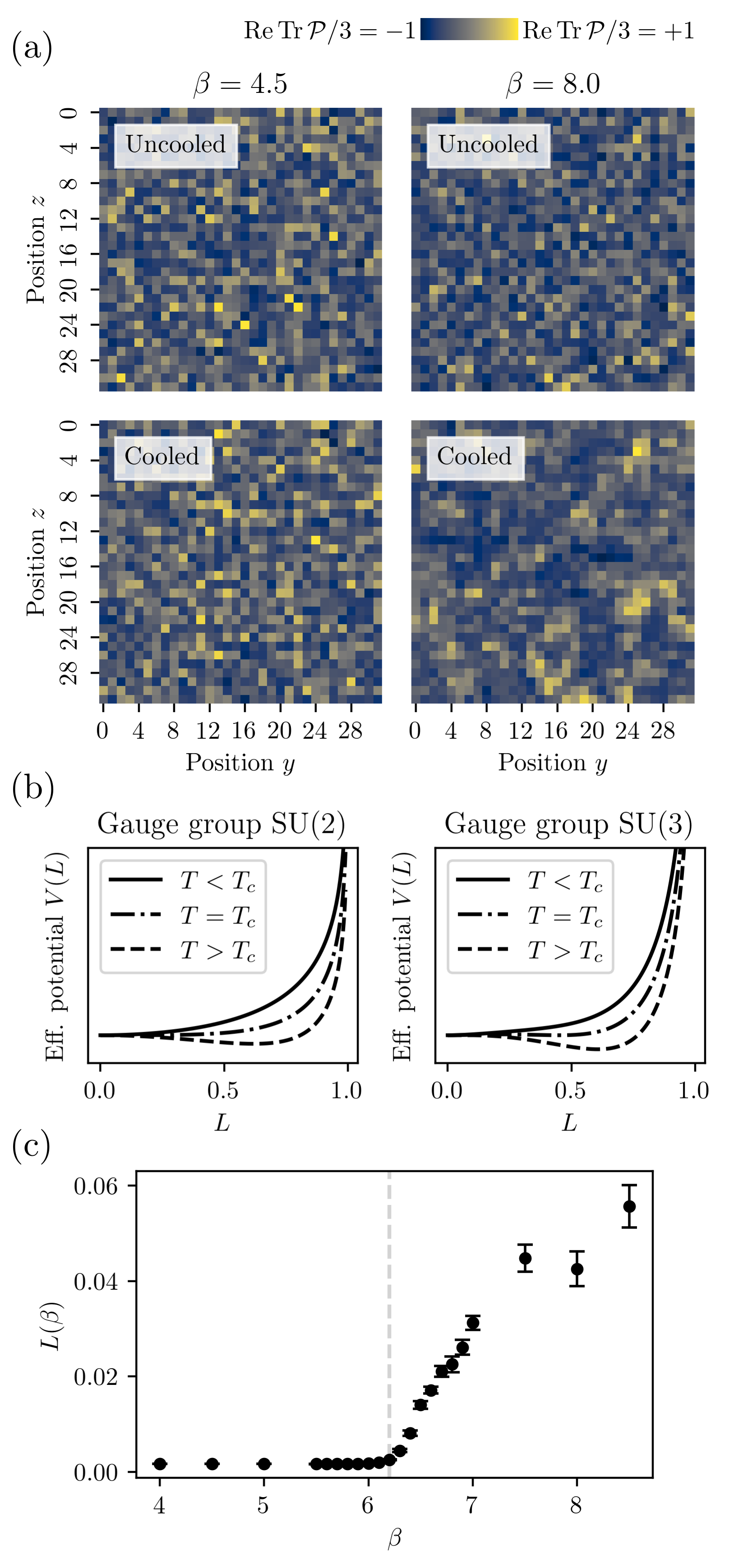}
	\caption{(a) Snapshots of the Polyakov loop trace $\RE\Tr \Pcal(\xx)/3$ at constant $\xx_1$ for $\beta<\beta_c$ (left) and $\beta>\beta_c$ (right), without cooling (top) and with cooling applied (bottom, $n_\flow = 20$).
 (b) Schematic effective Polyakov loop potentials in the confined phase (solid line), at the phase transition (dashed-dotted line) and in the deconfined phase (dashed line), derived from leading-order contributions of a strong coupling expansion and Haar measure contributions~\cite{Fukushima:2017csk}. 
 The left panel shows the potential for gauge group $\SU(2)$, which has a second-order phase transition, and the right panel shows the $\SU(3)$ potential, which exhibits a first-order phase transition.
 (c) Reference results for the volume average of the absolute Polyakov loop trace~($L$) in lattice units versus $\beta$ for uncooled data, where the vertical dashed line indicates $\beta_c\simeq 6.2$.}
    \label{Fig:PolyakovEffPotentialTrace}
\end{figure}

\subsection{First-order deconfinement phase transition}\label{Sec:FirstOrderDeconf}

An order parameter distinguishing between the confined and deconfined phases is given by the Polyakov loop, defined as the product of gauge links wrapping around the imaginary time direction,
\begin{equation}
    \Pcal(\xx) = \Pscr \prod_{\tau=1}^{N_\tau} U_4(\xx,\tau)\,,
\end{equation}
where $\Pscr$ denotes path ordering.
Some examples of two-dimensional slices of Polyakov loop trace configurations deep in the confined and deconfined phases are shown in \Cref{Fig:PolyakovEffPotentialTrace}(a), before and after cooling.
The cooling procedure reveals signals of extended structures for larger values of~$\beta$, whereas the configurations appear to remain largely disordered at low $\beta$.
The absolute volume average of the real part of the Polyakov loop trace,
\begin{equation}
    L = \frac{1}{3 N_\sigma^3} \bigg\langle \bigg| \sum_{\xx\in \Lambda_\sigma} \RE\Tr{\Pcal(\xx)}\bigg|\bigg\rangle\,,
\end{equation}
acquires a non-zero expectation value above some critical coupling marking the location of the phase transition, corresponding to a change of the shape of its effective potential; see \Cref{Fig:PolyakovEffPotentialTrace}(b).
For the particular setting chosen in the present work, the phase transition is located around $\beta_c\simeq 6.2$; see \Cref{Fig:PolyakovEffPotentialTrace}(c).\footnote{$\beta_c$ is provided here only as the approximate location where the Polyakov loop is observed to acquire a non-zero expectation value. A precise determination has been carried out in the literature~\cite{Boyd:1996bx} and is not the goal of the present work.}

The schematic effective Polyakov loop potentials shown in \Cref{Fig:PolyakovEffPotentialTrace}(b) highlight the key difference among the $\SU(2)$ and $\SU(3)$ phase transitions: the former is second and the latter is first order.
This is indicated by the different behavior of the location of the minimum around~$T_c$: for $\SU(2)$ it transitions from $L=0$ continuously to $L>0$, while for $\SU(3)$ a jump occurs at $T_c$.

Geometrically, a substantial difference between first and second-order phase transitions is a diverging correlation length for relevant excitations only in the second case.
Our previous study~\cite{Spitz:2022tul} has shown that a wide variety of geometric structures as detected by persistent homology can qualitatively change near second-order phase transitions in non-Abelian lattice gauge theories.
For first-order phase transitions, we therefore expect that fewer structures closely follow the phase transition dynamics with potentially less pronounced kinks at the critical temperature.
Instead, as we will reveal, there can be finite-volume effects, which resemble the behavior near second-order phase transitions but vanish in the infinite-volume effect.

\section{Duality signals in the Betti curves of electric and magnetic energy densities}\label{Sec:SelfDualExcitationsBettiElMagn}

TDA allows us to parametrize the landscape of minima, maxima and other critical points of functions on the lattice by means of extended topological structures along with measures of their dominance.
More specifically, sweeping through a sequence of nested topological spaces (called a filtration) inferred from the respective lattice function, their topologies can be efficiently described via homology.
Changes in the homology across the filtration are described by persistent homology, which we employ in this work and briefly introduce in \Cref{Sec:PersHomBackground}.

Powerful persistent homology-based observables are provided by the Betti curves, which count topological structures across the filtration.
In \Cref{Sec:ElMagnEnergyDensitiesBettiNumbers} we discuss them for local electric and magnetic energy densities.
The maximal number of topological structures present in these filtrations reveals hints towards the presence of electromagnetically dual excitations in the vicinity of~$\beta_c$, see \Cref{Sec:DualitySignals}.
We provide a tentative interpretation of these in light of well-known electromagnetic dualities.

\subsection{Background on persistent homology}\label{Sec:PersHomBackground}

We introduce the concept of persistent homology for sublevel sets of lattice functions, focusing on an intuitive approach.
We refer to the literature for comprehensive, mathematically more elaborate introductions~\cite{otter2017roadmap, ChazalMichelIntro}.

The sublevel sets of a real-valued function $f:\Lambda\to\rr$ on the lattice $\Lambda$ are given by
\begin{equation}
    M_f(\nu) := \{x\in \Lambda\,|\, f(x)\leq \nu\}
\end{equation}
for any $\nu \in \rr$.
For $\nu$ below $\min_{x\in \Lambda} f(x)$ the sublevel set is empty and for $\nu$ above $\max_{x\in\Lambda} f(x)$ the sublevel set is the entire lattice.
Furthermore, for $\nu \leq \mu$ we have $M_f(\nu)\subseteq M_f(\mu)$, so the family of sublevel sets $\{M_f(\nu)\}_\nu$ provides a \emph{filtration} of the lattice~$\Lambda$.

Yet, the sublevel sets $M_f(\nu)$ do not contain interesting topological information by themselves, since they merely are finite sets of lattice points.
Instead, one constructs so-called \emph{cubical complexes} from $f$, which resemble the sublevel sets $M_f(\nu)$ and are topologically less trivial.
Again, we provide a rather intuitive approach to their construction and refer to the literature for more elaborate treatments, see e.g.~\cite{kaczynski2004computational,wagner2012efficient}.
In general, a cubical complex is a set of cubes of different dimensions such as edges between points being dimension-1 cubes, squares being dimension-2 cubes and so forth, along with the requirement that the set be closed under taking boundaries.
The full cubical complex $\Ccal_\Lambda$ of the lattice $\Lambda$ consists of a 4-cube for each lattice point, where the lattice point is located at the cube's center.
It furthermore includes all 3-cubes, which appear as boundaries of the 4-cubes, all 2-cubes, which appear as boundaries of the 3-cubes, etc.
Finally, $\Ccal_\Lambda$ includes all vertices of the 4-cubes.

We use certain subsets of the full complex $\Ccal_\Lambda$ to describe the sublevel sets of the lattice function~$f$.
Specifically, we construct a function $\tilde{f}:\Ccal_\Lambda\to \rr$, whose sublevel sets provide subcomplexes of $\Ccal_\Lambda$, i.e., are again closed under taking boundaries.
First, for all 4-cubes $C\in\Ccal_\Lambda$ we set $\tilde{f}(C):=f(x)$, where $x\in \Lambda$ is the unique center point of $C$.
Any 3-cube $D\in\Ccal_\Lambda$ is contained in the boundary of two 4-cubes.
On the 3-cube $D$ the function $\tilde{f}$ picks up the lower of the two corresponding 4-cube values:
\begin{equation}\label{Eq:f3cube}
    \tilde{f}(D) := \min\{\tilde{f}(C)\,|\, D\in \partial C\}\,,
\end{equation}
which is repeated for all 3-cubes $D\in\Ccal_\Lambda$.
Analogously, all 2-cubes are contained in multiple 3-cubes, for which $\tilde{f}$ has been already defined.
\Cref{Eq:f3cube} can thus be consistently applied to all 2-cubes $D\in\Ccal_\Lambda$ for corresponding 3-cubes $C\in\Ccal_\Lambda$, similarly for the 1-cubes and the vertices.
This defines the function $\tilde{f}$ on all $\Ccal_\Lambda$.
Through the inductive construction, its sublevel sets
\begin{equation}
    \Ccal_f(\nu):=\{C\in \Ccal_\Lambda\,|\, \tilde{f}(C)\leq \nu\}
\end{equation}
form subcomplexes of the full cubical complex $\Ccal_\Lambda$.
For $\nu < \min_{x\in \Lambda} f(x)$: $\Ccal_f(\nu) = \emptyset$, while for $\nu\geq \max_{x\in\Lambda} f(x)$: $\Ccal_f(x) = \Ccal_\Lambda$.
Furthermore, for $\nu\leq \mu$ we have ${\Ccal_f(\nu)\subseteq \Ccal_f(\mu)}$, so the family $\{\Ccal_f(\nu)\}_\nu$ provides a filtration of the full cubical complex $\Ccal_\Lambda$.
This is called the \emph{sublevel set} or the \emph{lower-star filtration} of $f$, and $\nu$ is its filtration parameter.

\begin{figure}
    \centering
	\includegraphics[scale = 0.49]{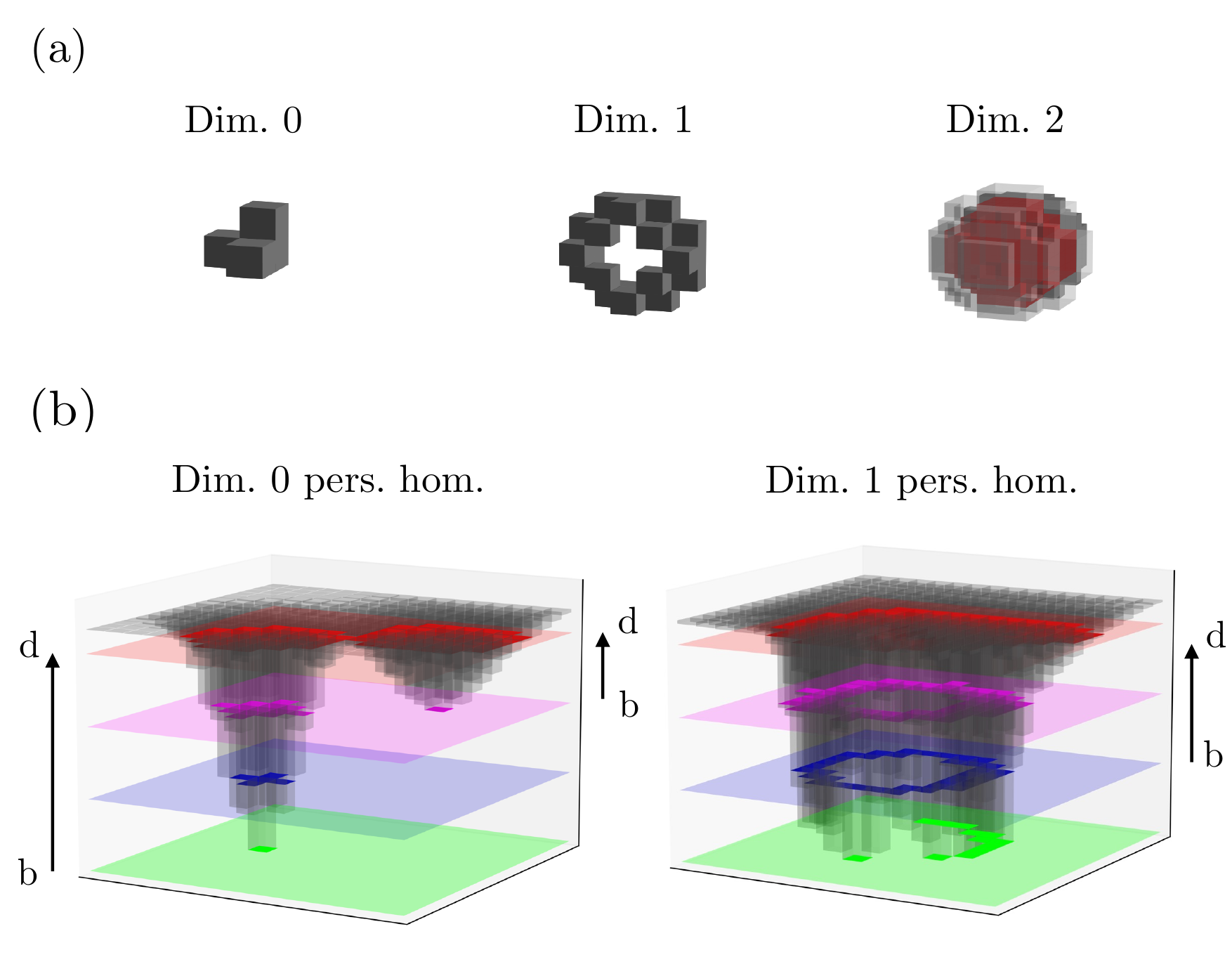}
	\caption{(a) Homology classes of different dimensions in a cubical complex; in dimension two red indicates an enclosed volume. 
 (b) Schematic illustration of persistent homology classes in a sublevel set filtration with birth and death parameters indicated, where exemplary sublevel set cubical complexes are depicted in different colors.
 The figure has been reprinted from~\cite{Noel:2023oyz} with the permission of the authors.}
    \label{Fig:PersHomIllustration}
\end{figure}

The cubical complexes $\Ccal_f(\nu)$ can be viewed as a `pixelization' of the sublevel sets $M_f(\nu)$ and are generally topologically less trivial than mere point sets.
Much of their topology is suitably described by means of homology~\cite{munkres2018elements}, which is algorithmically efficiently computable and homotopy invariant.
The $\Ccal_f(\nu)$ can give rise to non-trivial homology classes of different dimensions, as illustrated for three spatial dimensions in \Cref{Fig:PersHomIllustration}(a).
Indeed, different connected components and loop-like holes can appear, which form \mbox{dimension-0} respectively \mbox{dimension-1} homology classes.
Cubes can enclose empty volumes, which are described as \mbox{dimension-2} homology classes.
Finally, in four dimensions enclosed empty 4-volumes can appear, which provide \mbox{dimension-3} homology classes (not displayed).

Sweeping through the filtration $\{\Ccal_f(\nu)\}_\nu$, the homology may generally change depending on $\nu$.
This is illustrated in \Cref{Fig:PersHomIllustration}(b) for the case of functions $f$ defined on a two-dimensional lattice, where the vertical bar height indicates the function value.
Considering the left-hand example, when $\nu=\min_{x} f(x)$ (green plane), the first connected component (dimension-0 homology class) is born with birth parameter $b=\min_{x}f(x)$. 
Towards the larger filtration parameters indicated by the blue and purple planes, the single 2-cube evolves into a path-connected accumulation of 2-cubes.
The homology class corresponding to the first connected component remains invariant.
Yet, at the filtration parameter indicated by the purple plane, a second dimension-0 homology class is born.
At the filtration parameter indicated by the red plane, a saddle point occurs and the first homology class merges with the second.
The former dies with death parameter $d$.
The second homology class lives up to infinite filtration parameter; its death parameter $d$ can be formally set to $\infty$.

In addition to dimension-0 homology classes, dimension-1 homology classes may appear in the lower-star filtrations of functions on a two-dimensional lattice.
An example is provided on the right-hand side of \Cref{Fig:PersHomIllustration}(b), which resembles a vertically inverted `volcano'.
While for low filtration parameters as indicated by the green plane, multiple connected components appear, these all merge towards the larger filtration parameter indicated by the blue plane to form a loop-like hole.
A dimension-1 homology class has been born with birth parameter $b$.
Increasing the filtration parameter, it gets thickened (purple plane) and becomes ultimately fully filled with squares (red plane); it dies with corresponding death parameter $d$.

\begin{figure*}[t!]
    \centering
	\includegraphics[scale = 0.7]{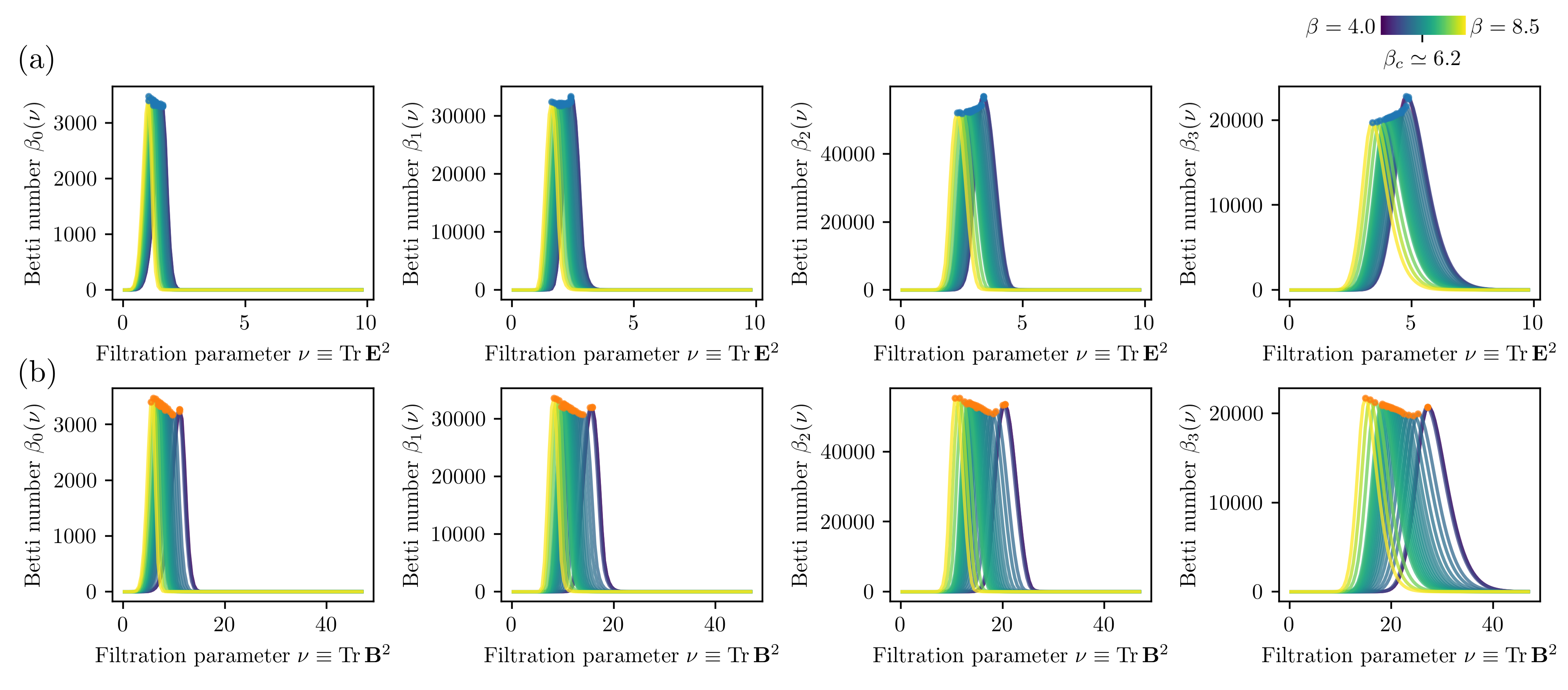}
	\caption{Betti curves for (a) the $\Tr{\EE^2}$ and (b) the $\Tr{\BB^2}$ sublevel set filtrations and for homology dimensions zero to three from left to right.
 Colors from blue to yellow correspond to $\beta$-values as indicated by the colorbar.
 The points located at the maxima highlight the maximal Betti curve values shown in \Cref{Fig:BettiCurvesEsqrBsqrMaxComparison}(a).}
    \label{Fig:BettiCurvesEsqrBsqr}
\end{figure*}

The \emph{persistent homology} of $\{\Ccal_f(\nu)\}_\nu$ is fully described by the collection of all birth-death parameter pairs $(b,d)$ for the homological features of the different dimensions.
The difference $p=d-b$ provides a measure for their dominance and is called \emph{persistence}.
In this work we mostly focus on the dimension-$\ell$ Betti curves, which count dimension-$\ell$ homology classes depending on the filtration parameter $\nu$:
\begin{equation}
    \beta_\ell(\nu) := \#\{\textrm{Dim.-}\ell\textrm{ homology classes of }\Ccal_f(\nu)\}\,.
\end{equation}

Persistent homology comes with a number of advantageous properties.
First, state-of-the-art algorithms allow for its efficient evaluation, computing the homology and related birth-death pairs for all filtration parameters at once.
We utilize the versatile computational topology library GUDHI for Python~\cite{boissonnat2014gudhi}.
It facilitates cubical complexes with periodic boundary conditions, which we employ.
Mathematically, persistent homology is provably stable with regard to perturbations of the input for a variety of persistent homology metrics, see e.g.~\cite{cohen2007stability, cohen2010lipschitz}.
This theoretically underpins its suitability for applications to lattice gauge theories.
Finally, persistent homology and the Betti curves can be well used in statistical analyses, giving rise to notions of ergodicity and large-volume asymptotics~\cite{hiraoka2018limit, spitz2020self}.

\subsection{Betti curves for electric and magnetic energy density filtrations}\label{Sec:ElMagnEnergyDensitiesBettiNumbers}

We turn to the Betti curves for the sublevel set filtrations of local electric and magnetic field energy densities, i.e., for ${f = \Tr\EE^2}$ and ${f=\Tr\BB^2}$.
For the lattice gauge theory under consideration, the total energy density reads 
\begin{equation}\label{Eq:TotalEnergyDensityProportionality}
T^{00}(\xx,\tau)\sim \Tr\EE^2(\xx,\tau)+\frac{1}{4}\Tr\BB^2(\xx,\tau)\,.
\end{equation}
Therefore, upon studying electric and magnetic energy density filtrations, we gain insights into the electromagnetic structures assembling the total energy density.
On the lattice, we employ clover-leaf variants of $\SU(3)$-valued electric and magnetic fields, which are provided by anti-symmetric combinations of spatio-temporal and spatial-only plaquettes, respectively.
Their construction has been outlined for the case of gauge group $\SU(2)$ in~\cite{Spitz:2022tul} and is not repeated here.
The prefactor $1/4$ for the magnetic contributions in \labelcref{Eq:TotalEnergyDensityProportionality} is due to the different normalization of the magnetic compared to the electric field.

\Cref{Fig:BettiCurvesEsqrBsqr} shows the Betti curves of homology dimensions zero to three for the $\Tr\EE^2$ filtration (panel (a)) and for the $\Tr\BB^2$ filtration (panel (b)), evaluated for a range of inverse couplings $\beta$.
All Betti curves provide sharply peaked distributions, which for increasing $\beta$-values shift towards lower filtration parameters.
The overall energy density decreases with increasing $\beta$, which explains this effect.

For increasing homology dimensions, the support of the distributions shifts towards larger filtration parameters and widens (from left to right).
This is a geometric effect due to the formation mechanism of homology classes of different dimensions.
For instance, multiple dimension-0 homology classes first need to be born in order to merge and form a dimension-1 feature, see also the right-hand example in \Cref{Fig:PersHomIllustration}(b).
Similarly, many dimension-1 features first need to form a pierced surface, which then gets successively filled with cubes to form an enclosed volume, i.e., a dimension-2 homology class.

\begin{figure*}[t]
    \centering
	\includegraphics[scale = 0.7]{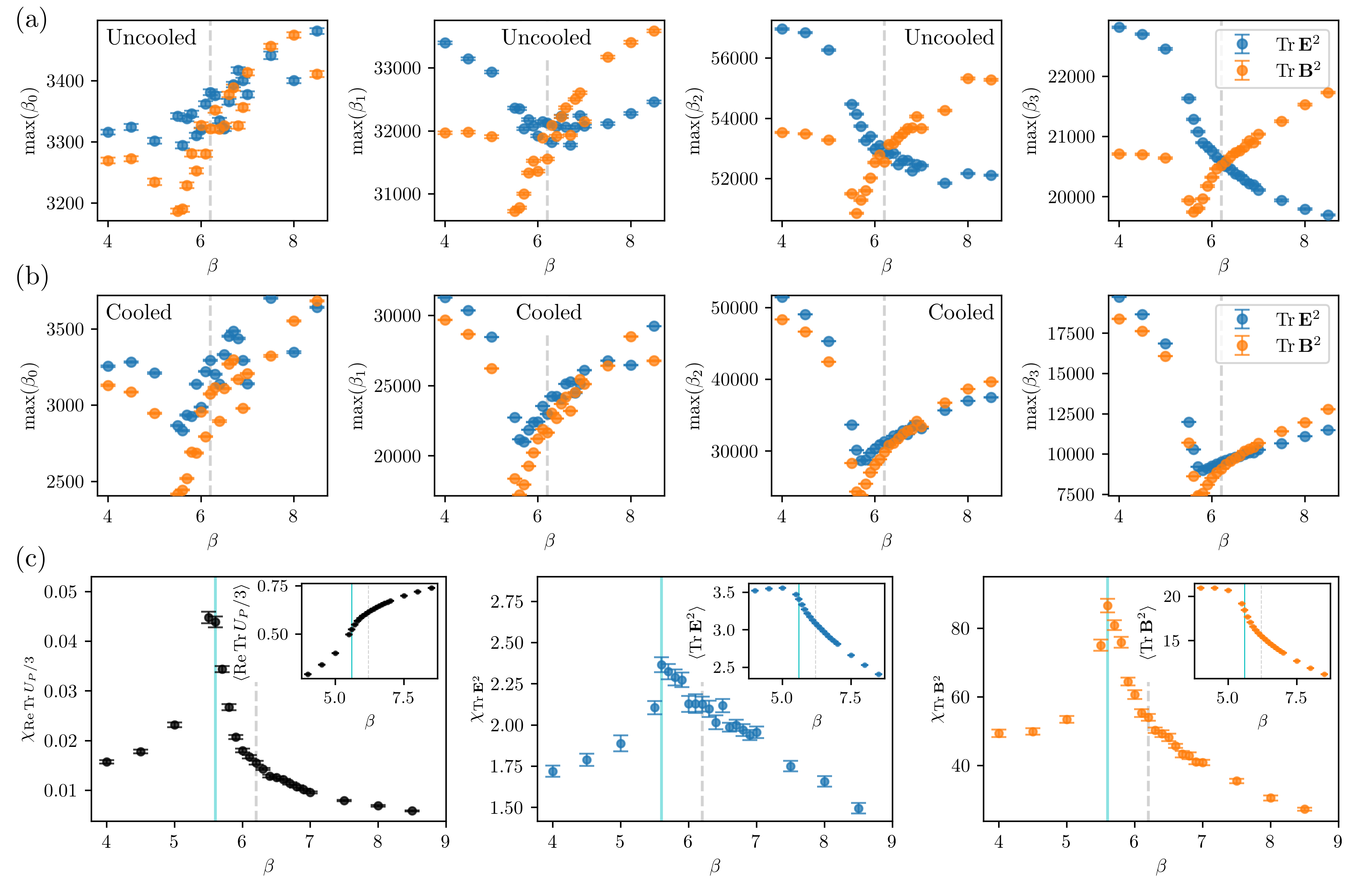}
	\caption{Maxima of the Betti curves given in \Cref{Fig:BettiCurvesEsqrBsqr} versus $\beta$ for homology dimensions zero to three from left to right, where results for the $\Tr{\EE^2}$ filtration are shown in blue and for the $\Tr{\BB^2}$ filtration in orange.
 (a) No cooling applied, (b) cooling applied with $n_\flow = 20$.
 (c) Reference results for the volume-scaled susceptibilities (as defined in \labelcref{Eq:Susceptibility}) of $\RE\Tr U_P/3$, $\Tr{\EE^2}$, and $\Tr{\BB^2}$ (from left to right) as a function of $\beta$, computed from the additional heatbath runs described in \Cref{Sec:LatticeSetup}.
 Insets show ensemble averages of the volume averages of these quantities themselves.
 The vertical light blue line indicates $\beta=5.6$, near which all susceptibilities reveal a distinct peak as a finite-volume effect.}
    \label{Fig:BettiCurvesEsqrBsqrMaxComparison}
\end{figure*}

Comparing the Betti curves for the $\Tr\EE^2$ and $\Tr\BB^2$ filtrations, we notice that in the latter case the support of the curves is at approximately a factor of 4 larger filtration parameters than in the former case.
This is due the different prefactor involved in the definition of $\BB$ compared to $\EE$, which also appears in the total energy density~\eqref{Eq:TotalEnergyDensityProportionality}.

For the $\Tr\EE^2$ filtration, the Betti curve peak height increases in homology dimension zero (left) with increasing $\beta$, but decreases for homology dimensions two and most strongly for three (right).
This is different for the $\Tr\BB^2$ filtration, where for all homology dimensions the peak heights, after a brief decline for low $\beta$, increase with increasing $\beta$.
We turn to a more detailed investigation of this phenomenon in the next subsection.

\subsection{Signals of electromagnetic dualities in Betti curve maxima}\label{Sec:DualitySignals}

In \Cref{Fig:BettiCurvesEsqrBsqrMaxComparison}(a) we show the peak values of the Betti curves of \Cref{Fig:BettiCurvesEsqrBsqr}, plotted against inverse coupling $\beta$.
The maximal value of a dimension-$\ell$ Betti curve is the maximal number of dimension-$\ell$ homology classes appearing in the filtration of cubical complexes.
\Cref{Fig:BettiCurvesEsqrBsqrMaxComparison}(b) shows the corresponding variables evaluated for the cooled/smoothed lattice configurations, and \Cref{App:CoolingImpact} provides a more detailed discussion on the dependence of the Betti curve maxima on the number of flow steps.
\Cref{Fig:BettiCurvesEsqrBsqrMaxComparison}(c) will be discussed later.

Naturally, the curves displayed in \Cref{Fig:BettiCurvesEsqrBsqrMaxComparison}(a) match the descriptions of the previous \Cref{Sec:ElMagnEnergyDensitiesBettiNumbers}.
Moreover, throughout all homology dimensions and for both uncooled and cooled configurations, the maximal Betti numbers for the $\Tr\BB^2$ filtration feature a local minimum near $\beta\simeq 5.6$.
The $\Tr\EE^2$ filtration has this effect only for the cooled configurations, for which, generally, the maximal Betti numbers are much closer to those of the $\Tr\BB^2$ filtration than for the uncooled case.
Throughout homology dimensions, the range of maximal Betti numbers differs for the uncooled configurations by up to $\sim 10\%$ within the displayed $\beta$-interval.
For the cooled configurations, the variation of the maximal Betti numbers depending on $\beta$ is of the order $\sim 40\%$ or more.

Most interestingly, at or near the critical $\beta_c\simeq 6.2$ we find a crossing among the maximal Betti numbers of the $\Tr\EE^2$ filtration with those of the $\Tr\BB^2$ filtration, in particular for homology dimensions one and larger.
This effect is most clearly visible in the top homology dimension three (right column).
Comparison among uncooled (panel (a)) and cooled (panel (b)) configurations shows that the crossing remains approximately stable against cooling, see also \Cref{Fig:BettiMaxEsqrBsqrCoolingComparison} in \Cref{App:CoolingImpact}.

We proceed with a discussion of possible interpretations of these findings.
We focus first on the crossings of maximal Betti numbers for the $\Tr \EE^2$ and $\Tr\BB^2$ filtrations at or near $\beta_c\simeq 6.2$, subsequently providing an explanation for the minima near $\beta\simeq 5.6$.
The crossings being most clearly visible in homology dimension three indicates that it is local maxima in electric and magnetic energy densities, which are predominantly responsible for these.
Indeed, unoccupied 4-cubes, which give rise to enclosed 4-volumes (dimension-3 homology classes), appear in the complexes $\Ccal_{\Tr \EE^2}(\nu)$ and $\Ccal_{\Tr\BB^2}(\nu)$ through local maxima in the corresponding lattice functions $\Tr\EE^2(\xx,\tau)$ and $\Tr\BB^2(\xx,\tau)$ for filtration parameters $\nu$ lower than these maximal values.
Therefore, the crossings hint at the presence of local lumps of field energy density switching type from electric to magnetic across the (de)confinement phase transition at $\beta_c$.

The crossings in the maximal Betti numbers for $\Tr\EE^2$ and $\Tr\BB^2$ filtrations are further reminiscent of electromagnetic dualities, such as semiclassically available for the Georgi-Glashow model~\cite{Montonen:1977sn,figueroa1998electromagnetic}.
Montonen-Olive dualities generally imply that the strong coupling behavior of the gauge theory can be determined by the dual theory at weak couplings, mapping the gauge bosons to magnetic monopoles and vice versa.
With electromagnetic dualities in mind, the crossings suggest the following interpretation.
Below $\beta_c$ and thus for bare couplings above the critical $g_c$, there is a larger abundance of homology classes, which correspond to local lumps of electric energy density, than there are lumps of magnetic energy density.
On top there are thermal fluctuations, which contribute a bit more to electric than to magnetic energy densities, cf. the $\Tr\EE^2$ filtration data in \Cref{Fig:BettiCurvesEsqrBsqrMaxComparison}(a) and (b).
Near the phase transition, electric and magnetic excitations contribute approximately equally to the structures appearing in the energy densities, resulting in the crossing of their maximal Betti numbers near $\beta_c$ and an equipartition in the number of electric and magnetic structures.
Above $\beta_c$ and therefore for bare couplings below the critical bare coupling $g_c$, the picture is reversed, so that electric structures get more scarce and magnetic structures in energy densities get more abundant.
Despite these considerations, we remind the reader that for the shown $\beta$-interval the variation of the maximal Betti numbers for both the $\Tr \EE^2$ and the $\Tr \BB^2$ filtration is of order $\sim 10\%$, so there appear to be significantly more structures present than those providing duality signals.

Cooling suppresses the thermal fluctuations overlaying the electromagnetic duality signatures, enhanced for electric excitations, see \Cref{Fig:BettiCurvesEsqrBsqrMaxComparison}(b).
Approaching self-duality, cooled configurations are closer to satisfying the Bogomol'nyi bound, so the structures in the $\Tr\EE^2$ and $\Tr\BB^2$ filtrations become more similar.
This is consistent with \Cref{Fig:BettiCurvesEsqrBsqrMaxComparison}(b).

The minima in the maximal Betti numbers near ${\beta\simeq 5.6}$ visible in \Cref{Fig:BettiCurvesEsqrBsqrMaxComparison}(a) and (b) can be understood as follows, based on susceptibilities.
The plaquette trace contributions to the action \eqref{Eq:Action} are given by
\begin{equation}\label{Eq:PlaquetteContrib}
    S_P[U] = \frac{1}{3} \sum_P \RE\Tr U_P\,.
\end{equation}
The $\beta$-derivative of the expectation value of \labelcref{Eq:PlaquetteContrib} is given by the plaquette susceptibility as defined in Eq.~\eqref{Eq:Susceptibility}:
\begin{align}
    \frac{\dd \langle S_P\rangle }{\dd \beta} = &\; \frac{\dd}{\dd\beta} \frac{\int \Dscr U \, S_P[U] \, \exp(\beta S_P[U])}{\int \Dscr U\, \exp(\beta S_P[U])}\nonumber\\
    =&\; \langle S_P^2\rangle - \langle S_P\rangle ^2 \equiv \frac{1}{V} \chi_{\RE\Tr U_P/3}\,,
\end{align}
where $\Dscr U$ is the lattice integral over all $\SU(3)$-valued link variables (constructed from $\SU(3)$ Haar measures).
The plaquette susceptibility $\chi_{\RE\Tr U_P/3}$ contains connected plaquette trace correlations which, in the vicinity of a second-order phase transition, are expected to converge to a non-zero constant at large separation.
This is due to a peak in the correlation length at the transition, for which the peak height grows to infinity in the infinite-volume limit.
The deconfinement phase transition of $\SU(3)$ being first order, no such diverging correlation length appears in our case.
Yet, as a finite-volume effect $\chi_{\RE\Tr U_P/3}$ peaks near $\beta\simeq 5.6$.
This can be inferred from the left panel in \Cref{Fig:BettiCurvesEsqrBsqrMaxComparison}(c), where $\chi_{\RE\Tr U_P/3}$ is shown as a function of $\beta$.%
~\footnote{For a lattice of size $12^4$, corresponding data has been discussed in~\cite{Gattringer:2010zz}, see Fig.~4.2 therein.}
This indicates that the structures, which contribute to the plaquette trace correlations, are largest around $\beta\simeq 5.6$, where their overall number must thus exhibit a minimum for the fixed lattice geometry.

Similar susceptibilities can be extracted from the local values of $\Tr\EE^2$ and $\Tr\BB^2$.
These are as well displayed in \Cref{Fig:BettiCurvesEsqrBsqrMaxComparison}(c).
We notice a distinct peak around $\beta\simeq 5.6$ in both susceptibilities, which is more pronounced for $\chi_{\Tr\BB^2}$ than for $\chi_{\Tr\EE^2}$.
This indicates that the correlation lengths of $\Tr\EE^2$ and $\Tr\BB^2$ excitations and therefore the related homology classes are largest near $\beta\simeq 5.6$.
The maximal Betti numbers thus exhibit a minimum around $\beta\simeq 5.6$, which we see in \Cref{Fig:BettiCurvesEsqrBsqrMaxComparison}(a) and (b).

The somewhat large variations observed in the maximal Betti numbers shown in \Cref{Fig:BettiCurvesEsqrBsqrMaxComparison}, in particular for homology dimensions zero and one, indicate that jackknife errors are likely underestimated, in particular when compared to the considerably larger uncertainties associated with the susceptibility results, despite the greater number of samples used there.
This is potentially due to larger than expected autocorrelation times of the Betti curve observables.
A detailed investigation of this issue is difficult due to the comparably high computational cost of the persistent homology analysis, and is beyond the scope of the present work.

\section{Robust features in Polyakov loop filtrations}\label{Sec:Polyakov}

An important order parameter for confinement is provided by the local Polyakov loop trace $P(\xx)$, described earlier in \Cref{Sec:FirstOrderDeconf}.
Moreover, many topological defects such as dyons can couple to Polyakov loops for a general gauge group $\SU(N_c)$, see e.g.~\cite{Diakonov:2007nv}.
This fact can be used to define a Polyakov loop-based variant of a topological density, whose integral over the spatial 3-torus yields the topological charge, see e.g.~\cite{Ford:1998mq}.

In \Cref{Sec:PersHomPolyakovTrace} we investigate the Betti curves for a filtration based on $P(\xx)$, giving rise to links to the previously discussed plaquette trace correlations.
\Cref{Sec:PersHomPolyakovTopDens} provides results for the Polyakov loop-based topological density filtration, which seems surprisingly robust against $\beta$-variations, at least if compared to the $\SU(2)$ case studied in~\cite{Spitz:2022tul}.

\subsection{Polyakov loop trace filtration}\label{Sec:PersHomPolyakovTrace}

\begin{figure*}[t]
    \centering
	\includegraphics[scale = 0.7]{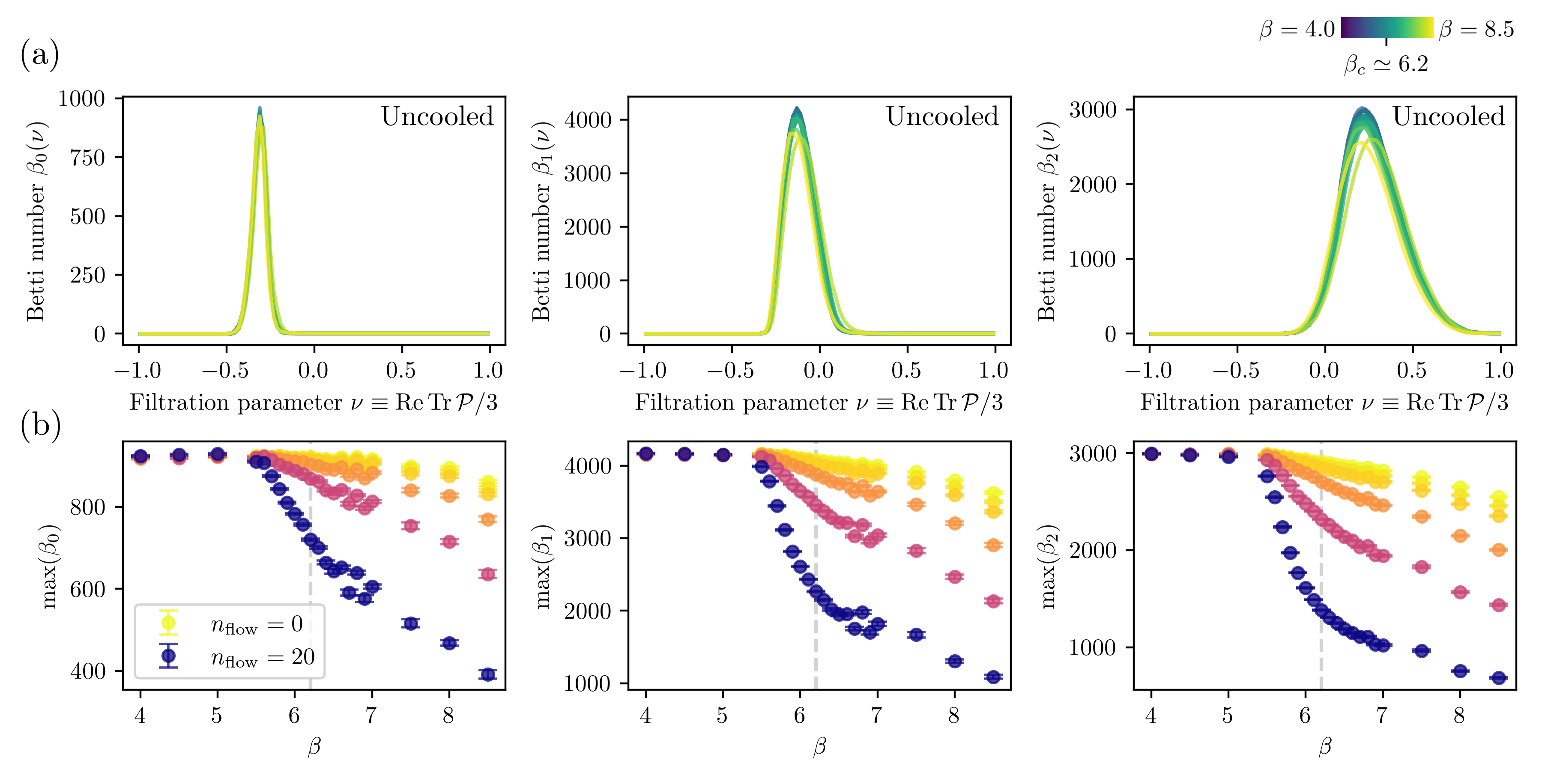}
	\caption{(a) Betti curves for the $\RE \Tr \Pcal/3$ filtration without cooling.
 (b) Corresponding maxima values plotted versus $\beta$, including a comparison with different flow steps ($n_\flow$) for cooling.
 The displayed flow steps are given by $n_\flow\in \{0,1,2,5,10,20\}$; shown colors interpolate between those printed in the legend.
 Homology dimensions zero to two are shown from left to right.
 The error bars shown in (b) have been computed via jackknife, see \Cref{App:JackknifeError}.}
    \label{Fig:BettiPolTr}
\end{figure*}

\Cref{Fig:BettiPolTr}(a) shows the Betti curves for the $\RE\Tr \Pcal/3$ filtration for a range of $\beta$-values, computed for uncooled configurations.
We find clearly peaked distributions of features across the filtration for all homology dimensions zero to two (left to right), whose support increases towards larger filtration parameters with increasing homology dimension.
Again, this is a natural finding for the sublevel set filtration, since multiple connected components first need to exist and to merge, in order to form a dimension-1 feature, and similarly for dimension-2 features.

In terms of their dependence on $\beta$, we notice that the Betti curves stay roughly on top of each other up to $\beta\simeq 5.6$, followed by a mild decline in peak heights for further increasing $\beta$-values.
This is more transparently visible in \Cref{Fig:BettiPolTr}(b), where the maximal values of the Betti curves of \Cref{Fig:BettiPolTr}(a) are displayed along with their dependence on the number of flow steps for cooling.
Clearly, the maximal Betti numbers stay approximately constant up to $\beta\simeq 5.6$, up to which value the peak heights remain also insensitive to cooling.
Above the kink around $\beta\simeq 5.6$, Betti curves begin to depend on the number of flow steps: increasing $n_\flow$ can strongly enhance the decline in peak heights with increasing $\beta$.

We encountered kink-like behavior around $\beta\simeq 5.6$ before: the maximal Betti numbers of $\Tr\EE^2$ and $\Tr\BB^2$, in particular for cooled configurations, gave rise to minima at this $\beta$-value, see \Cref{Fig:BettiCurvesEsqrBsqr} and the discussion towards the end of \Cref{Sec:DualitySignals}.
We attributed this behavior to a peak in plaquette trace correlations, which as a finite-volume effect occurs for our lattice near $\beta\simeq 5.6$ and not near $\beta_c \simeq 6.2$.
The plaquette trace correlations also correlating with the number of features in the Polyakov loop trace filtration, we expect the kink-like behavior in the related maximal Betti numbers near $\beta \simeq 5.6$ to also be a finite-volume effect.
The insensitivity of the maximal Betti numbers to cooling below $\beta\simeq 5.6$ can be attributed to the Polyakov loop behavior itself: at low $\beta$ cooling barely affects Polyakov loop traces and leaves the volume average zero, while at large $\beta$ cooling enhances the appearance of non-zero volume averages, cf. \Cref{Fig:PolyakovEffPotentialTrace}.
For $\beta>\beta_c$, thermal fluctuations on top of larger domains in $\RE\Tr \Pcal(\xx)$ get increasingly suppressed through cooling.

For gauge group $\SU(2)$ the Polyakov loop trace filtration has been studied in~\cite{Spitz:2022tul}.
It has been found that the Betti curves remain invariant to changes in $\beta$ up to the (pseudo-)critical $\beta_c$, above which Betti number distributions broaden and decrease in overall numbers for increasing $\beta$.
Qualitatively, this is similar to the $\SU(3)$ case investigated in the present work, except for the kink-like behavior occurring around $\beta\simeq 5.6$ and not near $\beta_c$, which, again, we expect to be a finite-volume effect.
For gauge group $\SU(3)$ this is possible, since the deconfinement phase transition is first order, while it is second order for gauge group $\SU(2)$.
Yet, the dependence of the Betti curves on $\beta$ above $\beta_c$ is much stronger in the $\SU(2)$ case than for $\SU(3)$ above $\beta\simeq 5.6$.
This can be explained at least partially via the Polyakov loop trace absolute volume average $L(\beta)$ having for $\SU(3)$ maximally $\sim 5\%$ of the value compared to $\SU(2)$, taking into account the studied $\beta$-intervals.
Therefore, it can be anticipated that for the given $\beta$-interval the $\beta$-dependence of the number of geometric structures associated with the Polyakov loop trace is weaker for $\SU(3)$ than for $\SU(2)$.

\subsection{Polyakov loop topological density filtration}\label{Sec:PersHomPolyakovTopDens}

\begin{figure*}[t]
    \centering
	\includegraphics[scale = 0.7]{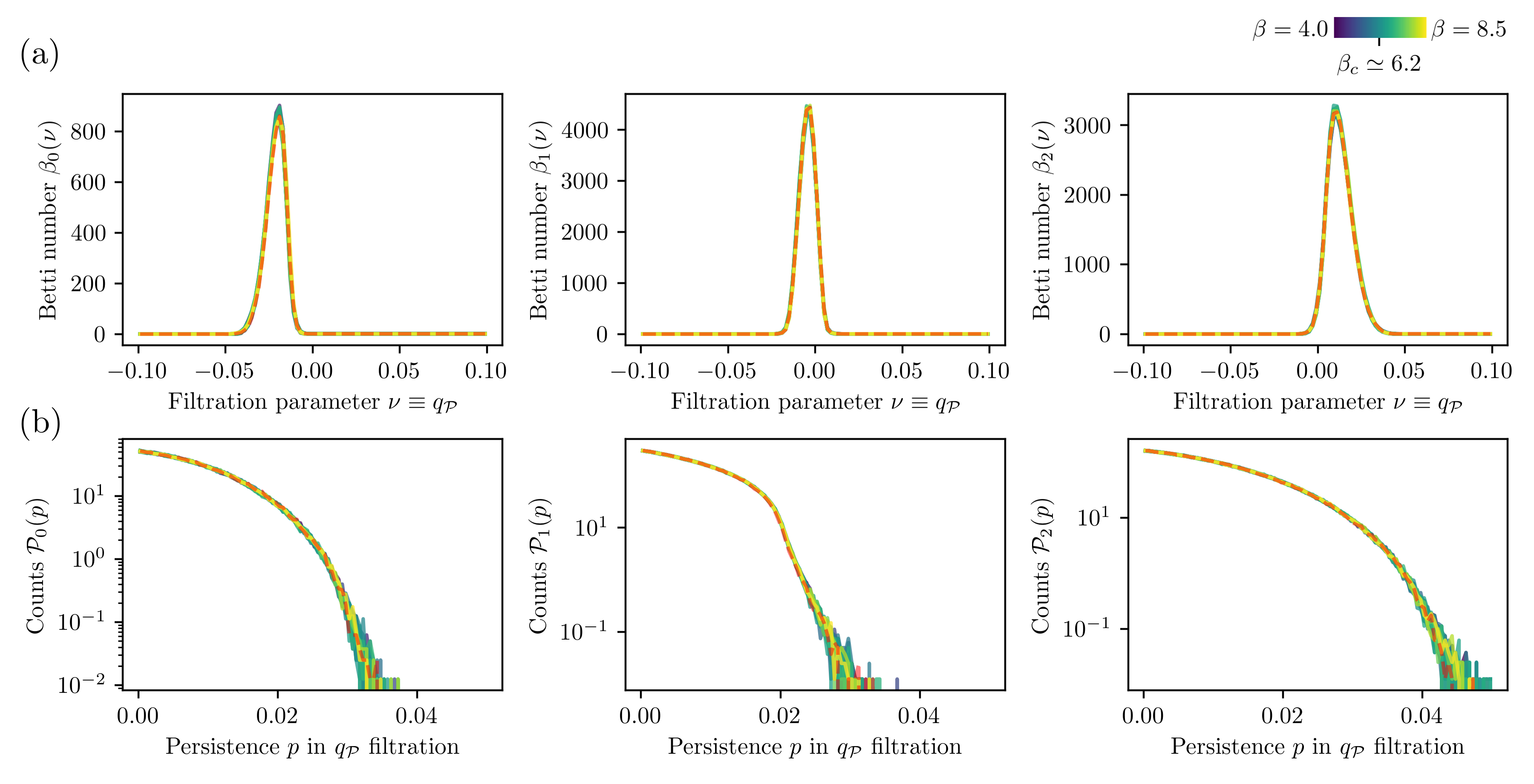}
	\caption{(a) Betti curves for the $q_\Pcal$ filtration.
 (b) Persistence distributions for the $q_\Pcal$ filtration. 
 Homology dimensions zero to two are shown from left to right.
 While the data for colors indicated by the colorbar have been computed without cooling, the red dashed lines indicate cooled data at $\beta=8.5$ for $n_\flow = 20$ flow steps.
 The results are both approximately $\beta$-insensitive and remain invariant under cooling.}
    \label{Fig:BettiPolTopDens}
\end{figure*}

We turn to investigate in how far topological excitations coupling to Polyakov loops might play a role for the dynamics visible in the $\Tr\EE^2$, $\Tr\BB^2$ and $\RE\Tr\Pcal/3$ filtrations.
For this we consider filtering the lattice configurations by means of the Polyakov loop topological density:
\begin{equation}
\begin{aligned}
    q_\Pcal(\xx) =&\; \frac{1}{32\pi^2} \varepsilon_{ijk}\Tr[(\Pcal^{-1}(\xx)\partial_i \Pcal(\xx))\nonumber\\
    &\qquad \times(\Pcal^{-1}(\xx)\partial_j \Pcal(\xx))(\Pcal^{-1}(\xx)\partial_k \Pcal(\xx))]\,,
\end{aligned}
\end{equation}
with $\varepsilon_{ijk}$ the Levi-Civita symbol in three dimensions.
Indeed, the nomenclature for $q_\Pcal$ is justified: for the pure gauge theory on a continuous space-time 4-torus, the topological charge can be computed as the integral of a topological density $\sim\Tr(\EE\cdot \BB)$ over the entire 4-torus.
The topological charge can be identically rewritten into an integral over the spatial 3-torus with integrand $q_\Pcal$, for gauge group $\SU(2)$~\cite{Ford:1998bt} as well as a general special unitary gauge group $\SU(N_c)$~\cite{Ford:1998mq}.
Similarly to our earlier work on persistent homology for gauge group $\SU(2)$~\cite{Spitz:2022tul}, it is expected that the persistent homology of the $q_\Pcal$ filtration is by construction less sensitive to lattice artifacts compared to a filtration using the usual topological density $\sim\Tr(\EE\cdot \BB)$.
This and the fruitful findings in~\cite{Spitz:2022tul} motivate the study of the $q_\Pcal$ filtration in the following.

In \Cref{Fig:BettiPolTopDens}(a) we show the Betti curves of the $q_\Pcal$ filtration, again for a range of $\beta$-values and mostly for uncooled configurations.
We find clearly peaked distributions, which barely reveal any $\beta$-dependence.
For all homology dimensions, the peak height does not reveal kink-like behavior but instead only randomly scatters by up to $\lesssim 4\%$ around mean values (maximal Betti numbers not displayed).
In \Cref{Fig:BettiPolTopDens}(a) the Betti curves for the uncooled configurations have been overlayed by the Betti curves for cooled configurations at $\beta = 8.5$, which agree with the corresponding data for the uncooled configurations.
This is consistent with the expected behavior of cooling: while it removes small-scale thermal fluctuations, it ideally leaves larger topological excitations untouched, therefore also the homological features appearing in the $q_\Pcal$ filtration.

The results for the Betti curves are complemented by the distributions of persistence values ($p=d-b$) for the $q_\Pcal$ filtration, shown in \Cref{Fig:BettiPolTopDens}(b).
Again, the curves with colors indicated by the colorbar have been computed for uncooled configurations, overlayed by the persistence distributions for cooled configurations at $\beta=8.5$.
We notice that all persistence distributions lay on top of each other, up to statistical fluctuations for the right-hand tail of the distributions.
Low statistics in this regime is responsible for these scatterings, since individual persistent homology classes and their persistences become visible.
Generally, the distributions for dimensions zero and two have similar shape, the latter coming with slightly larger support, while the dimension one persistence distributions look different.
This can be indicative for local values of $q_\Pcal$ scattering evenly around zero.
Indeed, if this is the case, then local minima in $q_\Pcal$ giving rise to signals in homology dimension zero and local maxima in $q_\Pcal$ dominating homology dimension two behave statistically alike, so do the related persistences.

Analogous to the Polyakov loop trace filtration discussed in \Cref{Sec:PersHomPolyakovTrace}, we have studied the Polyakov loop topological density filtration for gauge group $\SU(2)$ in~\cite{Spitz:2022tul}.
Interestingly, for $\SU(2)$ the persistence distributions of the $q_\Pcal$ filtration have revealed clear exponential behavior in homology dimensions zero and two.
We loosely attributed this to the presence of dyons, with fitted exponents heuristically matching the predictions for the topological charge statistics of dyons.
Furthermore, in the $\SU(2)$ case above $\beta_c$, cooling had a substantial influence on the persistence distributions, where it enhanced the presence of features with large persistences.
As we revealed, the situation for gauge group $\SU(3)$ is markedly different, since no exponential behavior occurs for the persistence distributions and cooling barely has any effect on the features occurring in the $q_\Pcal$ filtration.

\section{Conclusions}\label{Sec:Conclusions}

In the present work we have studied pure $\SU(3)$ lattice gauge theory on a Euclidean $32^3\times 12$ lattice through the lens of TDA.
The employed persistent homology of several observables has allowed for the extraction of robust homological features appearing in a variety of filtrations constructed from the field configurations.
We have focussed on filtrations based on local electric and magnetic energy densities as well as the real parts of Polyakov loop traces and a Polyakov loop-based topological density.

The Betti curves of electric and magnetic energy densities have proven interesting.
Considering the maximal numbers of homological features appearing across the filtrations (i.e., the maximal Betti numbers), we found new signals for electromagnetic dualities across the phase transition.
More specifically, below the phase transition local lumps of electric energy density dominate slightly over magnetic such lumps.
After an equipartition of the number of electric and magnetic features at the phase transition, the electric and magnetic behaviors interchange.
Even if Montonen-Olive dualities cannot be exactly realized in pure $\SU(3)$ gauge theory without supersymmetry, this raises the question if we nevertheless see at least partial indications for related excitations in the lattice gauge theory.

The maximal Betti numbers also included spatio-temporal geometric manifestations of plaquette trace correlations, which come with a finite-volume peak significantly below the phase transition.
Such behavior is in particular possible for first-order phase transitions but less likely for second-order transitions, and thus demonstrates that persistent homology can identify clear differences between phase transitions of first and second order.

Linking the results for the electric and magnetic energy density filtrations and the Polyakov loop-based filtrations, we notice that while qualitatively different behavior has occurred for the former two filtrations on both sides of the phase transition, this has not been the case for the Polyakov loop-based filtrations including the Polyakov loop topological density.
It is tempting to deduce from this that topological defects coupling to Polyakov loops may not be the only driving force behind the first-order deconfinement phase transition of pure $\SU(3)$ gauge theory, and neither is the formation of large domains in Polyakov loop traces, again based on the absence of qualitative changes.
Approximate electromagnetic dualities may also play a role in the transition, at least with regard to the persistent homology of the investigated filtrations.

A more detailed investigation of the nature of the related field configurations and their behavior across the phase transition is required.
Questions of interest are whether the relevant excitations can be described classically as solitons and how they relate to the known Montonen-Olive dualities.
Do they come with topological charges, and why is no such duality signal visible in the same filtrations for gauge group $\SU(2)$?
In this regard, also a detailed investigation of their local correlations with the Polyakov loop topological densities could be interesting.

We plan to extend the current topological data analysis to that of the dynamics of thermal phase transitions in QCD with dynamical fermions. The strongly correlated regime around and specifically above the pseudo-critical temperature $T_c$ is not fully resolved yet. Specifically, challenges concerning the temperature dependence of the axial anomaly, as well as that of the persistence and dynamics of topological correlations above $T_c$, persist, see e.g.~\cite{Sharma:2015wua, Mazur:2018pjw, Borsanyi:2022fub, Mickley:2024vkm}. Our analysis would add to the dissection of the strongly correlated analysis in this regime. We hope to report on this in the near future.

\begin{acknowledgments}

We thank J.~Berges, K.~Boguslavski, L.~de~Bruin, V.~Noel, P.E.~Shanahan, A.~Wienhard and A.~Wipf for discussions and work on related projects. 
This work is funded by the Deutsche Forschungsgemeinschaft (DFG, German Research Foundation) under Germany’s Excellence Strategy EXC 2181/1 - 390900948 (the Heidelberg STRUCTURES Excellence Cluster) and the Collaborative Research Centre, Project-ID No. 273811115, SFB 1225 ISOQUANT. 
JMU is supported in part by Simons Foundation grant 994314 (Simons Collaboration on Confinement and QCD Strings) and the U.S.\ Department of Energy, Office of Science, Office of Nuclear Physics, under grant Contract Number DE-SC0011090. 
This work is funded by the U.S.\ National Science Foundation under Cooperative Agreement PHY-2019786 (The NSF AI Institute for Artificial Intelligence and Fundamental Interactions, \url{http://iaifi.org/}).
\end{acknowledgments}

\appendix

\section{Impact of cooling on maximal Betti numbers for the $\Tr\EE^2$ and $\Tr\BB^2$ filtrations}\label{App:CoolingImpact}

\begin{figure*}[t]
    \centering
	\includegraphics[scale = 0.7]{Fig7_new.png}
	\caption{Maxima of the Betti curves versus $\beta$ for homology dimensions zero to three from left to right, where results for the $\Tr{\EE^2}$ filtration are shown in blue and for the $\Tr{\BB^2}$ filtration in orange.
    Rows (a) through (d) have been computed from samples with different numbers of flow steps $n_\flow = 1, 2, 5, 10$, respectively.
    Error bars have been computed via jackknife, see \Cref{App:JackknifeError}.
    This figure demonstrates the stability of the crossings among the $\Tr{\EE^2}$ and $\Tr{\BB^2}$ filtration Betti curve maxima against cooling.}\label{Fig:BettiMaxEsqrBsqrCoolingComparison}
\end{figure*}

In this appendix, we discuss the influence of cooling on the maximal Betti numbers for the $\Tr\EE^2$ and $\Tr\BB^2$ filtrations displayed in the main text, see \Cref{Fig:BettiCurvesEsqrBsqrMaxComparison}(a) to~(d).
For this, in \Cref{Fig:BettiMaxEsqrBsqrCoolingComparison} we display the maximal Betti numbers for a range of flow steps $n_\flow$ for cooling.
We notice that cooling has a strong influence on both the $\Tr\EE^2$ and the $\Tr\BB^2$ filtrations, whose maximal Betti numbers decrease in values for growing $n_\flow$.
This is natural for cooling: small-scale fluctuations are increasingly smoothed out with longer cooling times, so the overall number of features is expected to decrease.
Yet, the crossings among the maxima Betti numbers for the $\Tr{\EE^2}$ and $\Tr{\BB^2}$ filtrations remain stable against cooling, in particular for the top homology dimension.

For the $\Tr\EE^2$ filtration, without cooling no local minimum is present around $\beta\simeq 5.6$ in the maximal Betti numbers.
Yet, with increasing cooling times such a minimum starts to develop, mostly for $n_\flow \gtrsim 10$.
For the $\Tr\BB^2$ filtration, the shape of the maximal Betti numbers plotted against $\beta$ remains roughly insensitive to cooling with a local minimum at $\beta\simeq 5.6$ already there without cooling.
Overall, increasing cooling times result in an approach of the maximal Betti numbers for the $\Tr\EE^2$ and $\Tr\BB^2$ filtrations.
This might indicate the increasing dominance of self-dual excitations, the more cooling is applied.

\section{Uncertainty estimation for maximal Betti numbers via jackknife re-sampling}\label{App:JackknifeError}

In this appendix, we describe how the uncertainties on the maximal Betti numbers $\max(\beta_\ell)$ for the $\Tr\EE^2$, $\Tr\BB^2$ and $\RE\Tr\Pcal/3$ sublevel set filtrations are computed.
We estimate errors via the statistical jackknife procedure as outlined e.g.~in~\cite{Gattringer:2010zz}.
Let $b_i := \max(\beta_\ell)^{(i)}$ be the maximal Betti number for any one of the filtrations, evaluated for the $i$-th sample of in total $N$ samples.
The (biased) mean is defined as
\begin{equation}
    \hat{b}:=\frac{1}{N} \sum_{i=1}^N b_i\,.
\end{equation}
We construct $N$ subsets of the original sample index set $\{1,\ldots,N\}$ by removing the $i$-th sample.
The means of these samples are denoted $\hat{b}_i$, where entry $b_i$ has been removed, accordingly.
We then define the variance
\begin{equation}
    \sigma^2:=\frac{N-1}{N} \sum_{i=1}^N \big(\hat{b}_i - \hat{b}\big)^2\,,
\end{equation}
whose square root provides an estimate for the standard deviation of $\hat{b}$.
Throughout this work, we show maximal Betti numbers including their uncertainties as $\hat{b}\pm \sigma$.

\section{Maximal Betti numbers for gauge group $\SU(2)$}\label{App:MaxBettiNumbersSU2}

\begin{figure*}[t]
    \centering
	\includegraphics[scale = 0.7]{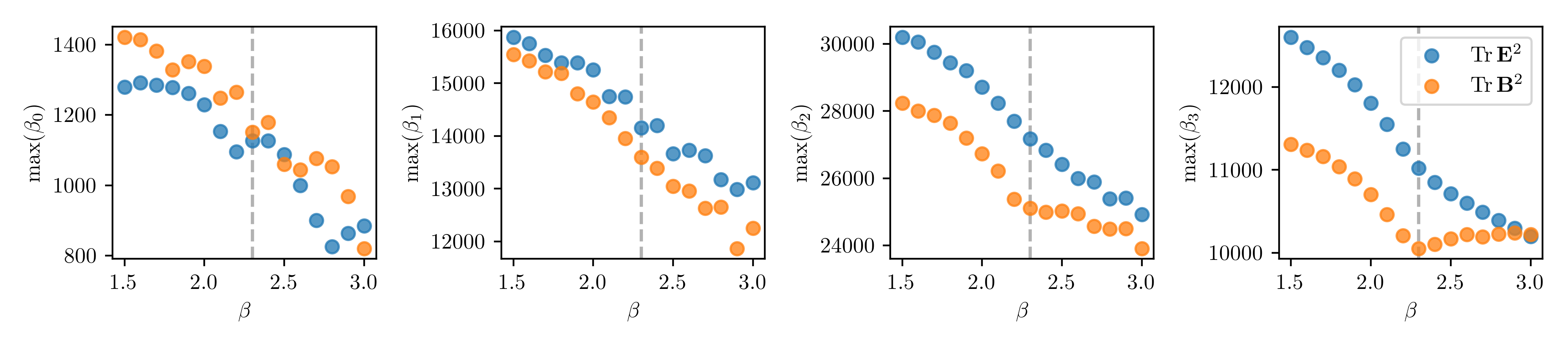}
	\caption{Maxima of Betti curves for lattice gauge theory with gauge group $\SU(2)$ on a $32^3\times 8$ lattice, shown as a function of $\beta$ for homology dimensions zero to three from left to right.
 Results for the $\Tr{\EE^2}$ filtration are shown in blue and for the $\Tr{\BB^2}$ filtration in orange.
 No cooling has been applied.
 The data of~\cite{Spitz:2022tul}, published by the same authors as the present work, has been reevaluated for the generation of this figure.}\label{Fig:BettiCurvesEsqrBsqrMaxComparisonSU2}
\end{figure*}

For comparison with the results of \Cref{Sec:SelfDualExcitationsBettiElMagn}, in this appendix we discuss the maximal Betti numbers of the $\Tr\EE^2$ and $\Tr\BB^2$ filtrations for gauge group $\SU(2)$.
In~\cite{Spitz:2022tul} we elaborated on the corresponding Betti curves with the crucial difference of having computed the \emph{superlevel set filtration}, not the sublevel set filtration.
For the former, function values above a certain threshold are of relevance, not below as for the sublevel sets.
Accordingly, we reevaluated our $\SU(2)$ configurations to compute the maximal Betti numbers of the $\Tr\EE^2$ and $\Tr\BB^2$ sublevel set filtrations for a range of $\beta$-values ranging from $1.5$ to $3.0$.

The results are shown for the uncooled configurations in \Cref{Fig:BettiCurvesEsqrBsqrMaxComparisonSU2}.
In~\cite{Spitz:2022tul} we have identified $\beta_c=2.3$ as the pseudo-critical inverse coupling, which has been highlighted by the dashed vertical line in the figure.
Mostly, we find monotonously decreasing curves.
We notice that in homology dimensions zero and one no qualitative change occurs near $\beta_c$.
This is different for homology dimension two, where the $\Tr\BB^2$ filtration reveals kink-like behavior near $\beta_c$.
Most clearly, the phase transition is visible in homology dimension three, where the curve for the $\Tr\EE^2$ filtration changes type from concave to nearly linear behavior.
The $\Tr\BB^2$ filtration data decreases up to $\beta_c$, then exhibits a minimum at $\beta_c$, and subsequently increases again.

Considering gauge group $\SU(3)$, the major finding of the present work has been crossings in the maximal Betti numbers for the two filtrations exactly at $\beta_c$, which approximately remained insensitive to cooling and has been present for all homology dimensions (see \Cref{Sec:DualitySignals}).
For gauge group $\SU(2)$, no such crossings are visible, except for a crossing in homology dimension three around $\beta\simeq 3.0$.
To conclude this appendix, the maximal Betti numbers for the $\Tr\EE^2$ and $\Tr\BB^2$ filtrations for gauge groups $\SU(2)$ and $\SU(3)$ are significantly different.

\nocite{spitz_2025_15256813}

\bibliography{literature}

\end{document}